\documentclass[lettersize,journal]{IEEEtran}

\usepackage{setspace,cite}

\usepackage{multirow,multicol}
\usepackage[table]{xcolor}

\usepackage[tbtags]{amsmath}
\usepackage{amsbsy}
\usepackage{amssymb}
\usepackage{amsfonts}
\usepackage{amsthm}

\usepackage{graphicx}
\usepackage{algorithmic}
\usepackage{algorithm}
\usepackage{balance}
\usepackage{rotating}
\usepackage{multirow}
\usepackage{subcaption}
\usepackage{fancyvrb}
\usepackage{latexsym}
\usepackage{verbatim}
\usepackage{esint}

\usepackage[pagebackref=true,breaklinks=true,letterpaper=true,colorlinks,bookmarks=false]{hyperref}

{\begin{list}               
    {$\bullet$ \hfill}{
        \setlength{\leftmargin}{\parindent}
        \setlength{\parsep}{0.04\baselineskip}
        \setlength{\itemsep}{0.5\parsep}
        \setlength{\labelwidth}{\leftmargin}
        \setlength{\labelsep}{0em}}
    }
{\end{list}}

\providecommand{\eref}[1]{Equation~(\ref{#1})}  
\providecommand{\cref}[1]{Chapter~\ref{#1}}

\providecommand{\fref}[1]{Figure~\ref{#1}}

\providecommand{\E}{\ensuremath{\mathbb{E}}}

\renewcommand{\vec}[1]{\ensuremath{\boldsymbol{#1}}}
\providecommand{\mat}[1]{\ensuremath{\boldsymbol{#1}}}

\providecommand{\calD}{\mathcal{D}}

\providecommand{\calF}{\mathcal{F}}

\providecommand{\mA}{\mathbf{A}}

\providecommand{\mI}{\mathbf{I}}

\providecommand{\mL}{\mathbf{L}}

\providecommand{\mS}{\mathbf{S}}

\providecommand{\mU}{\mathbf{U}}

\providecommand{\va}{\mathbf{a}}

\providecommand{\ve}{\mathbf{e}}
\providecommand{\vf}{\mathbf{f}}

\providecommand{\vr}{\mathbf{r}}
\providecommand{\vs}{\mathbf{s}}
\providecommand{\vt}{\mathbf{t}}
\providecommand{\vu}{\mathbf{u}}

\providecommand{\vx}{\mathbf{x}}
\providecommand{\vy}{\mathbf{y}}
\providecommand{\vz}{\mathbf{z}}

\providecommand{\mSigma}{\mat{\Sigma}}

\providecommand{\valpha}{\vec{\alpha}}
\providecommand{\vbeta}{\vec{\beta}}

\providecommand{\vtheta}{\vec{\theta}}

\providecommand{\vxi}{\vec{\xi}}

\providecommand{\vrho}{\vec{\rho}}

\providecommand{\vvarphi}{\vec{\varphi}}

\newcommand{\trace}[1]{\mathop{\mathrm{Tr}\left(#1\right)}}

\begin{document}

\title{Real-Time Dense Field Phase-to-Space Simulation of Imaging through Atmospheric Turbulence}

\author{Nicholas~Chimitt,~\IEEEmembership{Student~Member,~IEEE,}
        Xingguang~Zhang,~\IEEEmembership{Student~Member,~IEEE,}
        Zhiyuan~Mao,~\IEEEmembership{Student~Member,~IEEE,}
        Stanley~H.~Chan,~\IEEEmembership{Senior~Member,~IEEE}
\thanks{The authors are with the School of Electrical and Computer Engineering, Purdue University, West Lafayette,
IN, 47907 USA. Email: \{nchimitt, zhan3275, mao114, stanchan\}@purdue.edu.}
\thanks{The research is based upon work supported in part by the Intelligence Advanced Research Projects Activity (IARPA) under Contract No. 2022‐21102100004, and in part by the National Science Foundation under the grants CCSS-2030570 and IIS-2133032. The views and conclusions contained herein are those of the authors and should not be interpreted as necessarily representing the official policies, either expressed or implied, of IARPA, or the U.S. Government. The U.S. Government is authorized to reproduce and distribute reprints for governmental purposes notwithstanding any copyright annotation therein.}
}

\maketitle

\begin{abstract}
Numerical simulation of atmospheric turbulence is one of the biggest bottlenecks in developing computational techniques for solving the inverse problem in long-range imaging. The classical split-step method is based upon numerical wave propagation which splits the propagation path into many segments and propagates every pixel in each segment individually via the Fresnel integral. This repeated evaluation becomes increasingly time-consuming for larger images. As a result, the split-step simulation is often done only on a sparse grid of points followed by an interpolation to the other pixels. Even so, the computation is expensive for real-time applications. In this paper, we present a new simulation method that enables \emph{real-time} processing over a \emph{dense} grid of points. Building upon the recently developed multi-aperture model and the phase-to-space transform, we overcome the memory bottleneck in drawing random samples from the Zernike correlation tensor. We show that the cross-correlation of the Zernike modes has an insignificant contribution to the statistics of the random samples. By approximating these cross-correlation blocks in the Zernike tensor, we restore the homogeneity of the tensor which then enables Fourier-based random sampling. On a $512\times512$ image, the new simulator achieves 0.025 seconds per frame over a dense field. On a $3840 \times 2160$ image which would have taken 13 hours to simulate using the split-step method, the new simulator can run at approximately 60 seconds per frame.
\end{abstract}

\begin{IEEEkeywords}
Atmospheric turbulence, wave propagation, Zernike basis, Phase-to-Space Transform, Fourier optics
\end{IEEEkeywords}

%
\IEEEpeerreviewmaketitle

\section{Introduction}
\begin{figure}[th]
    \centering
    \includegraphics[width=\linewidth]{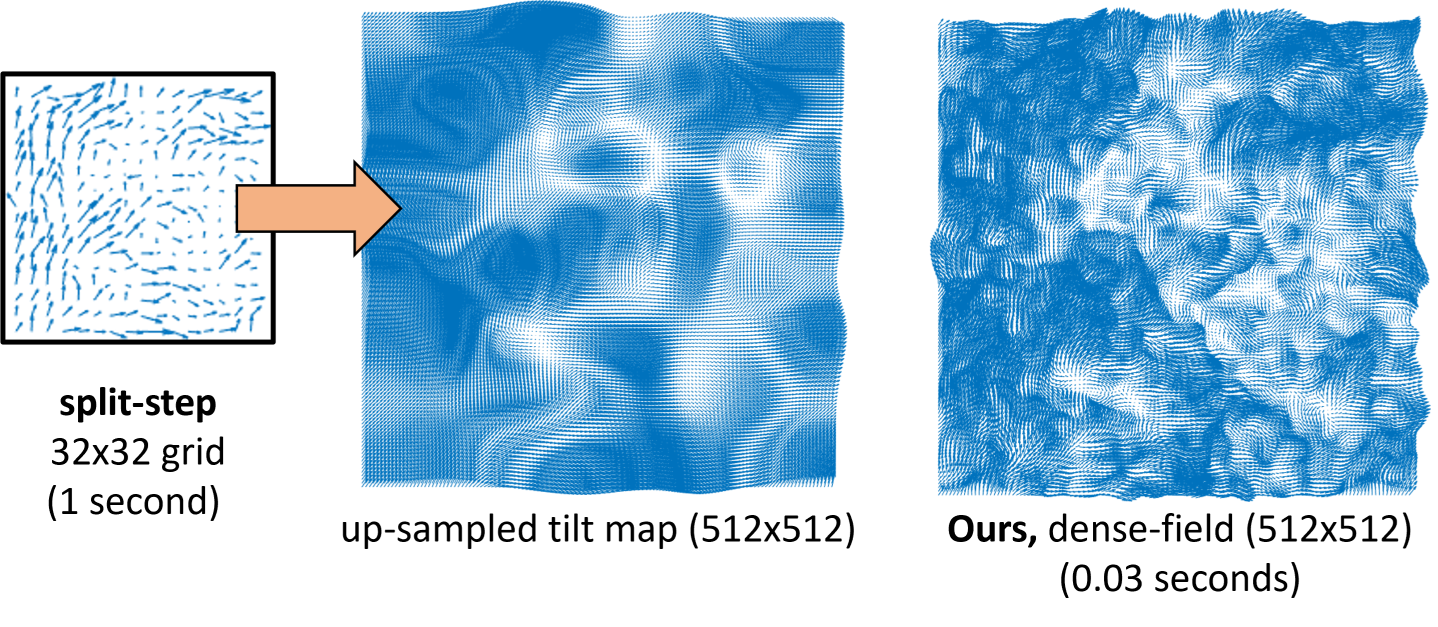}
    \includegraphics[width=\linewidth]{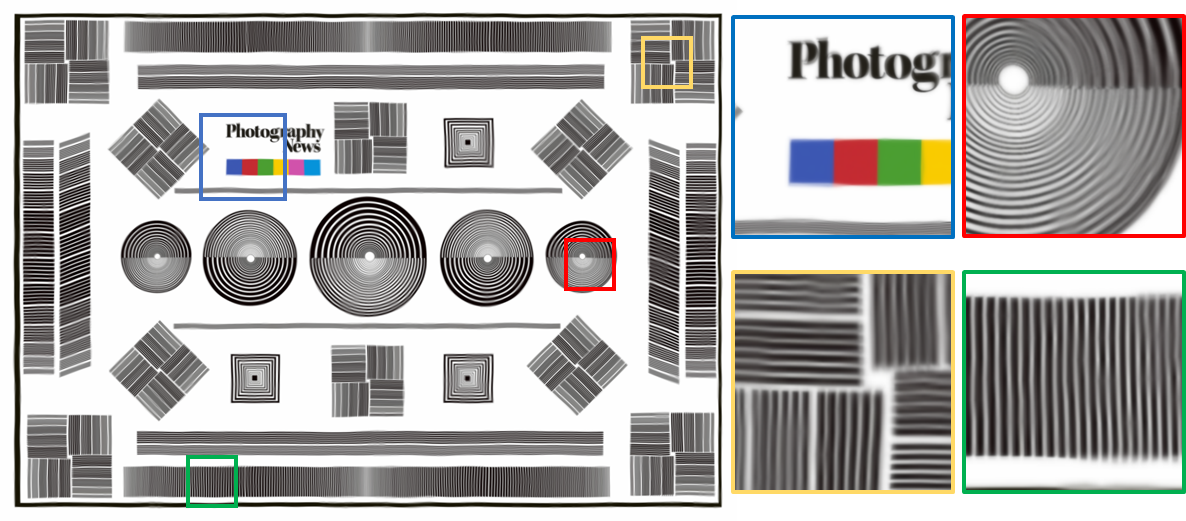}
    \caption{[Top] This paper presents a new turbulence simulation method that produces \emph{dense-field} turbulence in real-time. For a $512\times512$ image, the classical split-step propagation takes 1 second to generate a $32 \times 32$ grid followed by interpolation of the field. The proposed method takes 0.03 seconds to generate the turbulence at dense-grid. [Bottom] Snapshot of a simulated turbulence image from an input image with a 4K resolution ($3840\times2160$ pixels). }
    \label{fig: Fig01 overview}
\end{figure}

Light propagating through the atmosphere suffers from distortions due to the random spatio-temporal fluctuations in the index of refraction. Over a long distance, these distortions will accumulate and degrade the image quality. The development of atmospheric turbulence mitigation algorithms has received a considerable amount of interest over the past few decades \cite{Anantrasirichai2013, RoggemannFrameSelection, Anan2018, Gilles2016, Hardie2017, Milanfar2013, mao_tci, Mao_2022_ECCV, Chan_2022_TiltBlur}. Deep-learning-based techniques have recently been reported with some promising preliminary results \cite{Vpatel_single_image_turb, chellapa_restoration, patel_confidence, Zhang_2022_TMT}. However, as in any inverse problem, the atmospheric turbulence problem requires a forward model that can accurately describe the image formation process. To this end, simulating the turbulent effect becomes a critical step toward the goal of designing algorithms, evaluating methods, and understanding the limitations of imaging systems.

For decades, simulating atmospheric turbulence is most accurately performed in the wave domain because there is no simple intensity domain model. The ``gold standard'' approach is the split-step propagation \cite{RoggemannSimulator, Hardie2017, SchmidtTurbBook}.\footnote{There exist other modalities for the simulation of these effects that do not require computationally costly numerical wave propagation \cite{voronstov1_sim, voronstov1_sim, rucci_sim_sparse}, however, split-step is presently the most theoretically justifiable approach.}  The idea is to split the propagation path into segments and model the phase distortion in each segment for every pixel individually. However, split-step propagation is not scalable. For a $256 \times 256$ sized image, the split-step simulator reported in \cite{Hardie2017} can evaluate a grid of size $64 \times 64$ where the remaining pixels are interpolated. The reported runtime was approximately 24.6 seconds per frame on a GPU. If the size of the image grows to $3840 \times 2160$ (4K resolution) and the grid is \emph{dense}, i.e. without interpolation, a rough estimate is about 13.7 hours for one image. To synthesize a training dataset containing 1000 of these sequences where each sequence has 100 frames, this will take 156 years. Recognizing the pressing need for an accurate and fast turbulence simulator, we present a method that enables turbulence simulation in real time.

A preview of our results is shown in \fref{fig: Fig01 overview}. While split-step propagation is largely limited to a small grid of sprase points, the proposed method can directly generate a \emph{dense field}. The run-time of the proposed method is approximately 0.025 seconds for a $512 \times 512$ image. For a high-definition (HD) image of size $3840 \times 2160$, the runtime is approximately 60 seconds. As can be seen in \fref{fig: Fig01 overview}, the new simulator allows us to zoom in to any region of the image while the turbulent effect is still globally correlated according to the theoretical statistics. To our knowledge, this is the first practical demonstration of an HD dense-field turbulence simulation documented in the literature.

The proposed approach, named the Dense Field Phase-to-Space (DF-P2S) simulation, is built upon the multi-aperture model by Chimitt and Chan \cite{Chimitt2020}, and the phase-to-space (P2S) transform by Mao et al. \cite{Mao_2021_ICCV}. DF-P2S overcomes a fundamental limitation of \cite{Mao_2021_ICCV} which is the size of the cross-correlation matrix (tensor). In \cite{Mao_2021_ICCV}, the cross-correlation tensor must be pre-computed, stored, and decomposed before running the simulator. This causes some computational overhead, however, the bigger issue is memory. The largest cross-correlation tensor that can be stored is for a spatial grid of size $32\times32$ using 36 basis coefficients. To simulate an image with a higher resolution, we need to interpolate the field, which limits the overall accuracy of such a simulation. Our solution to overcome this memory bottleneck is to maintain the homogeneity along the spatial dimensions of the correlation tensor and perform an approximation of the cross-correlation functions which otherwise restrict this behavior. This is based on a new observation that the exact form of the cross-correlation functions can be approximated without severely hurting the tensor statistics. As a result, we can employ Fourier-based techniques to draw \emph{dense field} samples spatially at a low computational cost. This allows us to maintain a similar speed as P2S \cite{Mao_2021_ICCV}, yet gain an increase in statistical accuracy.

To summarize, this paper offers two contributions:
\begin{enumerate}
\item \textbf{Real-time dense field turbulence simulation.} We report the first turbulence simulator that can simulate over a dense field and in real-time.
\item \textbf{New approximation to the cross-correlation function.} We show how certain off-diagonal blocks of the cross-correlation matrix used in \cite{Mao_2021_ICCV} can be removed to utilize Fourier-based generation, hence enabling a significant resolution upscale.
\end{enumerate}

\section{Outline of General Simulation Principles: Building Blocks}
To keep track of the notations, we use object plane coordinates $\vx = (x,y)$ and image plane coordinates $\vu = (u,v)$. For a function defined across the aperture, we use the coordinates $\vxi = (\xi,\eta)$ or its polar form $\vrho = (\rho,\theta)$. We also use the polar coordinates $\vs = (s, \varphi)$ to denote the displacement in the Zernike space.

The turbulence effect is modeled by convolving a \emph{spatially varying} point spread function (PSF) to a diffraction-limited clean image. In the case of incoherent light, which is the focus of this work, the observed image $I(\vx)$ is
\begin{equation}
    I(\vx) = h_{\vx} (\vu) \circledast I_g(\vu),
    \label{eq: spatially_varying_convolution}
\end{equation}
where $h_{\vx}(\vu)$ is the PSF with the subscript to emphasize that it is spatially varying, and $I_g(\vu)$ is the distortion-free image. Note that the observed image is indexed by $\vx$ whereas the PSF and ideal image are indexed by $\vu$. This is to emphasize that after $h_{\vx}(\vu)$ is convolved with $I_g(\vu)$, only the \emph{center pixel} is used to construct $I(\vx)$.

The per-pixel PSF can be generated per the Fraunhofer diffraction equation with phase error \cite{Goodman_FourierOptics}. Denoting $P(\vxi)$ as the aperture function, $h_{\vx}(\vu)$ is
\begin{equation}
    h_{\vx}(\vu) = \left|\mathfrak{Fourier}\left\{ P(\vxi) e^{-j\phi_{\vx}(\vxi)} \right\}\right|^2,
    \label{eq: PSF_formation}
\end{equation}
withholding some constants that determine the size of the PSF according to the optical parameters. Here, $\phi_{\vx}(\vxi)$ is the phase distortion function that varies over $\vxi$ for coordinate $\vx$ in the image. Note that $\phi_{\vx_0}(\vxi) \neq \phi_{\vx_1}(\vxi)$ if $\vx_0 \not= \vx_1$.

\begin{figure*}[th]
    \centering
    \includegraphics[width=0.9\linewidth]{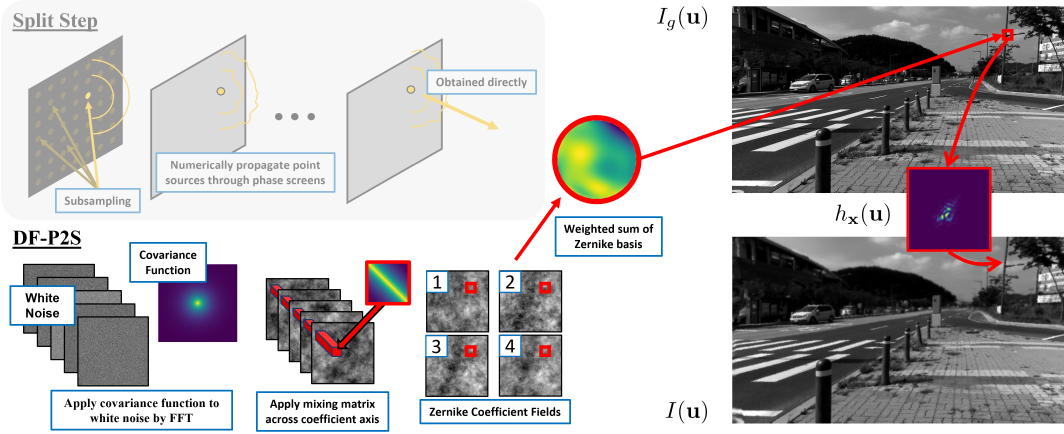}
    \caption{Here we show the main difference between classical approaches of modeling the medium and propagation directly, \cite{RoggemannSimulator, Hardie2017}, and our approach based on \cite{Chimitt2020, Mao_2021_ICCV}. In split-step, a subset of pixels in the image have point sources propagated through the medium which are then interpolated between to form the image at full resolution. In our approach, every single point has its own basis vector representation, meaning a phase function is generated for each pixel \emph{without} interpolation. We emphasize that both approaches generate phase realizations for the image formation process, but it is the generation approach that differs.}
    \label{fig: split_vs_multi}
\end{figure*}

Given that the PSF generation \eqref{eq: PSF_formation} and the image formation \eqref{eq: spatially_varying_convolution} is relatively standard, the central focus of a simulation approach then falls upon the generation of the random phase $\phi_{\vx}(\vxi)$ in accordance with its theoretically given statistics. There are two main categories for generating $\phi_{\vx}$: (i) split-step propagation \cite{RoggemannSimulator, Hardie2017, SchmidtTurbBook}, which numerically propagates a wave through a random volume, and hence \emph{modeling the medium}; (ii) collapsed phase-over-aperture \cite{Chimitt2020}, which generates the phase function directly at the aperture, which we refer to as the \emph{multi-aperture model}. We illustrate the differences in these two approaches in \fref{fig: split_vs_multi}.

\subsection{Computational Bottleneck of Split-Step}
Before we discuss the two building blocks of our simulator, it would be useful to highlight the limitations of the split-step method \cite{RoggemannSimulator, Hardie2017, SchmidtTurbBook}. The split-step method directly mirrors the physical process by which light propagates. After a point propagates through the simulated medium, it arrives at the aperture of the imaging system with a phase component $\phi(\vxi)$. In the case of turbulence, the statistics of $\phi(\vxi)$ is determined by the structure function
\begin{equation}
    \calD_{\phi}(\vxi, \vxi') = \E[(\phi(\vxi)-\phi(\vxi'))^2].
\end{equation}
Assuming that the random function $\phi(\vxi)$ is homogeneous and isotropic, the structure function can be simplified to
\begin{equation}
    \calD_{\phi}(|\vxi - \vxi'|) = 6.88(|\vxi - \vxi'|/r_0)^{5/3},
    \label{eq: fried_struct}
\end{equation}
where $r_0$ is the Fried parameter \cite{Fried66optical}.

To numerically generate the phase $\phi$, the split-step method uses the Kolmogorov power spectrum density (PSD) \cite{Tatarski_1967_a} (or similarly Von Karman spectrum \cite{Hardie2017}, etc) to generate discrete planes of turbulent distortions, referred to as phase screens \cite{SchmidtTurbBook}. This can be done directly with knowledge of the PSD and the Fourier transform based random field generation approach. The phase screens are placed along the propagation path as shown in \fref{fig: split_vs_multi}. The size of the phase screens is generated larger than the input image in order to model the spatial correlations which are determined by the overlap of the phase screen along the path.

Since the split-step models the medium directly, it is regarded as the most theoretically justifiable approach. A comprehensive discussion of this model is provided in \cite{SchmidtTurbBook}. However, the main limitation of split-step is its computational requirements. For every point source, we need to perform multiple fast Fourier transforms (FFTs). A simulation with $M$ phase screens results in $M (W \times H)$ 2D FFTs for an $W \times H$ image, with each FFT being at the size of the image. Scaling the process of generating a large dataset is nearly impossible.

\subsection{Building Block 1: Multi-Aperture Model}
Recognizing the speed limit of the split-step simulation, Chimitt and Chan proposed a new concept in \cite{Chimitt2020} which they named the \emph{multi-aperture model}. The idea is to skip the propagation by going to the statistical description of the resultant phase $\phi_{\vx}(\vxi)$ using the Zernike representation first proposed by Noll in 1976 \cite{Noll_1976}. The idea is to define a radius $R$ and a vector $\vrho$ such that $R\vrho$ is the polar coordinate representation of $\vxi$. Then, the phase $\phi_{\vx}$ can be represented as
\begin{equation}
\underset{\phi_{\vx}(\vxi)}{\underbrace{\phi_{\vx}(R\vrho)}} = \sum_{j=1}^{\infty} a_{\vx,j} Z_j(\vrho) \approx \sum_{j=1}^{N} a_{\vx,j} Z_j(\vrho),
\label{eq: Zernike decompose}
\end{equation}
with $Z_j(\vrho)$ as the $j$th Zernike function and $a_{\vx,j}$ as its respective coefficient. We emphasize that the phase $\phi_{\vx}$ is location specific, i.e., the phase function at $\vx$ is different from the phase at $\vx'$ if $\vx \not= \vx'$. Therefore, the Zernike coefficients $a_{\vx,j}$ are different from $a_{\vx',j}$. However, the basis function $Z_j(\vrho)$ is shared for all locations. In this paper, we set $N=36$, though this can easily be increased at the cost of some speed.

The coefficients $a_{\vx,j}$ are zero-mean Gaussian, and have a correlation matrix (technically, a tensor stored in the matrix form) which we denote the $(\vx,\vx',i,j)$th element of the matrix $\mA$ as
\begin{equation}
    [\mA]_{\vx,\vx',i,j} = \E [a_{\vx,i} \; a_{\vx',j}].
    \label{eq: A matrix}
\end{equation}
This notation stresses that the matrix $\mA$ is location dependent on the pair $\vx$ and $\vx'$. In the special case when the correlation matrix is spatially invarying, we write $[\mA]_{\vx,\vx',i,j}$ as $[\mA]_{\vx-\vx',i,j}$. If we further assume that $\vx' = \vx$, the correlation matrix becomes $[\mA]_{\vec{0},i,j} = \E [a_{\vx,i} \; a_{\vx,j}]$, which is equivalent to the covariance matrix given by Noll \cite{Noll_1976}.

Based on the decomposition in \eref{eq: Zernike decompose}, we can simulate the turbulence in three steps \cite{Chimitt2020}: (i) Collapse the screens; (ii) Draw spatially correlated Zernike vectors; (iii) Draw inter-modally correlated Zernike vector. Therefore, we have effectively converted the split-step propagation into a sampling problem of the Zernike coefficients.

This multi-aperture model provides roughly a $6\times$ increase in speed over split-step. However, the major drawback is that the spatial correlation for the higher order ($a_{\vx,i}$ where $i \ge 3$) Zernike terms are not computationally feasible. The generation of the spatial and intermodal correlated vectors requires a large covariance matrix, which exponentially increases in size with the size of the image. Therefore, only tilts are reasonably implementable for this method. Furthermore, the image formation process is the same as split-step, requiring us to generate one PSF for one pixel repeatedly over the entire field of view plus convolutions to produce the final image.

\begin{figure*}[th]
    \centering
    \includegraphics[width=\linewidth]{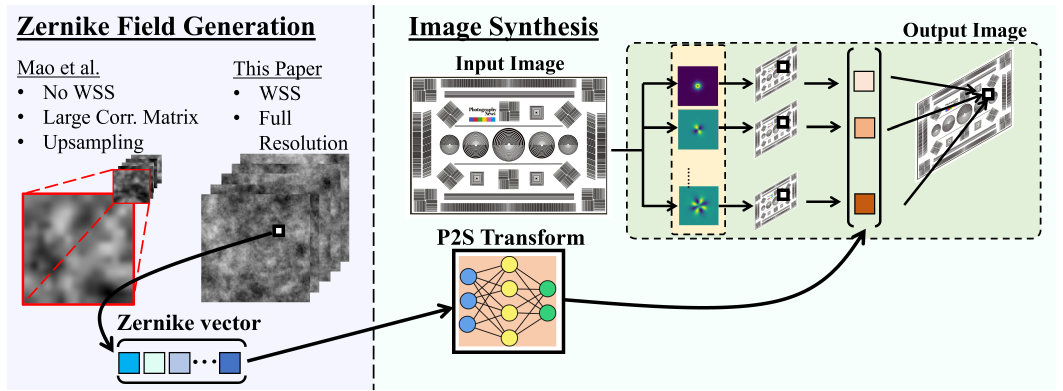}
    \caption{Here we present the overall DF-P2S simulation pipeline. We first generate a Zernike vector for \emph{every} point in the image, a feat only possible through our simulation technique. Next, we feed these Zernike vectors parallely into the P2S transform. The P2S transform maps the phase coefficients to PSF coefficients, which are then used to implement a fast version of a spatially varying convolution.}
    \label{fig: dfp2s}
\end{figure*}

\subsection{Building Block 2: Phase-To-Space}
While the multi-aperture approach reduced the computation by collapsing the propagation into a sampling problem, the PSFs still need to be generated according to \eqref{eq: PSF_formation} and the spatially varying convolution still needs to be performed via \eqref{eq: spatially_varying_convolution}. To address these limits in speed, Mao et al. \cite{Mao_2021_ICCV} introduced the phase-to-space (P2S) transform  that reformulates the PSF via a \emph{spatial} decomposition:
\begin{equation}
    h_{\vx}(\vu) = \sum_{m=1}^M \beta_{\vx,m} \vvarphi_m(\vu),
    \label{eq: p2s PSF}
\end{equation}
where $\vvarphi_m(\vu)$ is a \emph{spatial} basis function for the PSF. In \cite{Mao_2021_ICCV}, these basis functions are learned by running the principle component analysis on a database on PSFs. The parameter $\beta_{\vx,m}$ denotes the coefficient associated with each basis $\vvarphi_m(\vu)$. What \eref{eq: p2s PSF} says is that the spatially varying convolution in \eqref{eq: spatially_varying_convolution} can now be replaced by a set of spatially invariant convolutions using $\vvarphi_m(\vu)$. To form the image, we simply compute the per-pixel coefficient $\beta_{\vx,m}$ and form the weighted average:
\begin{align}
    I(\vx)
    &= h_{\vx}(\vu) \circledast I_g(\vu) \notag \\
    &= \sum_{m=1}^M \beta_{\vx,m} \underset{\text{spatially invariant convolution}}{\underbrace{\vvarphi_m(\vu) \circledast I_g(\vu)}},
    \label{eq: varying blur}
\end{align}
where we should emphasize the speed improvement due to the spatially invariant convolution. However, the biggest challenge here is that there is no analytic expression that allows us to translate the Zernike basis $Z_j(\vrho)$ to the spatial basis $\vvarphi_m(\vu)$. The solution proposed in Mao et al. \cite{Mao_2021_ICCV} is to train a small neural network to convert the coefficients from the Zernike coefficients $\va_{\vx} = [a_{\vx,1}, a_{\vx,2}, \ldots, a_{\vx,N}]$ (which are sampled from $\mA$) to the PSF coefficients $\vbeta_{\vx} = [\beta_{\vx,1}, \beta_{\vx,2}, \ldots, \beta_{\vx,M}]$.

The phase-to-space transform eliminates the necessity of taking any FFT for the formation of a PSF, dramatically speeding up the generation process by $1000\times$ compared to split-step. However, the limitation of the phase-to-space transform is the memory requirement. For an image of size $W \times H$, the generation using $36$ Zernike coefficients requires the construction of a matrix that is $36HW \times 36HW$. What is worse is that we can not leverage the homogeneity property (analogous to the wide-sense stationary in random processes) and so we cannot use Fourier transforms to decompose the correlation matrix and draw random samples. The workaround solution in \cite{Mao_2021_ICCV} is to focus on a small grid of points $H/d \times W/d$ where $d$ is the sampling ratio, followed by decomposing the $36HW/d^2 \times 36HW/d^2$ matrix and interpolating between these points. The grid sizes, in most of the demonstrations, are in the range of $16 \times 16$ to $32 \times 32$ due to memory limitations. In this paper, we resolve this memory issue by completely eliminating the interpolation, thus enabling a truly dense field turbulence generation.

\subsection{Other Simulation Approaches}

While split-step is regarded as the most accurate simulation method, other approaches do exist. The most closely related one is the brightness function simulation \cite{voronstov1_sim, Vorontsov_2005_a}. These methods are based on the solution to ray optics simulation through a series of phase screens. Their comparisons with the split-step can be found in \cite{Lachinova2017}. Additionally, empirically-driven methods such as Hunt et al. \cite{rucci_sim_sparse} model the atmospheric distortion basis representations. There are also low-order approximation methods such as \cite{Milanfar2013, Lau2017, Leonard_Howe_Oxford}, which can generate visually plausible images but fail to satisfy the turbulence statistics.

\section{Dense Field Phase-to-Space Simulation (DF-P2S Simulation)}

In this section, we outline the proposed DF-P2S simulation. DF-P2S uses the same backbone as the original P2S in \cite{Mao_2021_ICCV} but is more efficient in terms of memory. DF-P2S enables a dense field simulation that was not possible in P2S.

The overall simulation can be broken down into two steps as shown in \fref{fig: dfp2s}:
\begin{enumerate}
    \item \textbf{Generate correlated Zernike coefficients.} The goal of this step is to use Building Block 1: Multi-Aperture Model to draw random but correlated Zernike coefficients according to the correlation matrix.
    \item \textbf{Apply the Zernike coefficients using the P2S transform.} This step uses Building Block 2: Phase-to-Space Transform to convert the correlated Zernike coefficients to the spatial basis coefficients. Consequently, the resulting image can be formed by \eref{eq: varying blur}.
\end{enumerate}

The biggest innovation of DF-P2S is a new way to accomplish the first step. We ask the question: how can we efficiently draw the Zernike coefficients from $\mA$ without suffering from the memory bottleneck? To answer this question, it is important to investigate the structure of the correlation matrix $\mA$.

\subsection{Intuition of the New Method}
The idea of drawing correlated Zernike coefficients according to a correlation matrix is not difficult if the number of grid points is small. Consider a multivariate Gaussian random variable with mean $\E[\vy] = 0$ and the correlation matrix $\mSigma = \E[\vy\vy^T]$. To draw a random vector $\vy$ from this distribution, we can decompose the correlation matrix $\mSigma = \mU\mS\mU^T$ via the eigen-decomposition, and define $\mSigma^{\frac{1}{2}} = \mU\mS^{\frac{1}{2}}\mU^{T}$. Then, starting with a white noise vector $\ve \sim \text{Gaussian}(0,\mI)$, the transformed vector $\vy = \mSigma^{\frac{1}{2}} \ve$ will satisfy the desired property that $\E[\vy] = 0$ and $\E[\vy\vy^T] = \mSigma$.

When the size of the grid is small so that the dimension of the correlation matrix $\mSigma$ is small, the matrix $\mSigma^{\frac{1}{2}}$ can be generated using standard numerical techniques such as the Cholesky factorization. However, for a large grid of points, storing the matrix $\mSigma$ and running the factorization would become infeasible. One exception is that if $\vy$ is homogeneous (wide-sense stationary), then the correlation matrix $\mSigma$ is \emph{circulant} and so the eigen-decomposition is equivalent to the Fourier transform. In this case, generating the random vector $\vy$ can be implemented via
\begin{equation*}
    \vy = \mU\mS^{\frac{1}{2}}\mU^{T} \ve = \calF^{-1}( \mS^{\frac{1}{2}} \calF(\ve) ),
\end{equation*}
where $\calF$ denotes the discrete-time Fourier transform, and the diagonal matrix $\mS$ is the Fourier spectrum of one row (or column) of the correlation matrix $\mSigma$.

The significance of the homogeneity is that it allows us to speed up the sampling process by performing all computations in the Fourier space. In addition, the memory bottleneck is resolved because we do need to construct the full correlation matrix $\mSigma$ and run the Cholesky factorization. The question is: For the Zernike correlation matrix we are considering in this paper, does it have any kind of wide sense stationarity? If not, can we approximate it using something that has such a property?

\subsection{Structure of the Correlation Matrix}
\begin{figure}[h]
    \centering
    \includegraphics[width=0.95\linewidth]{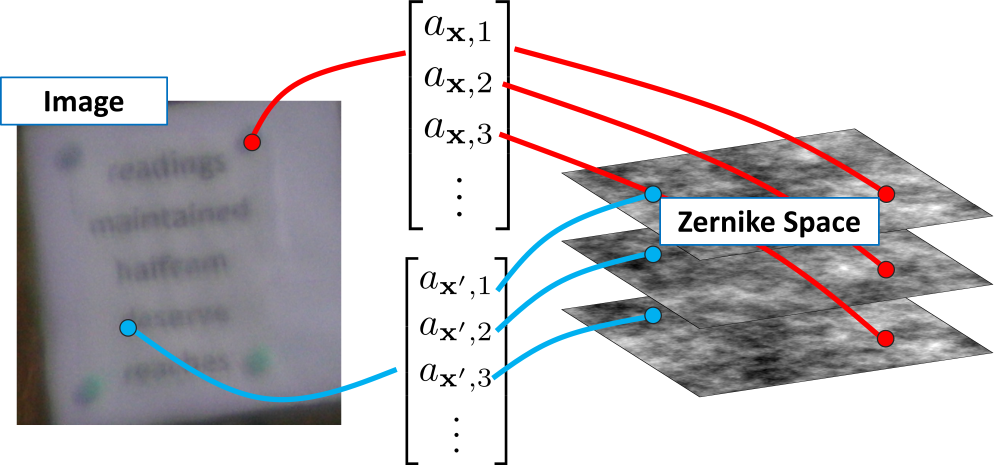}
    \caption{The Zernike Space is a representation describing the phase decomposition by its basis functions. For every point on an object, or analogously every pixel in an image, there is a Zernike vector that describes the distortion across the aperture. This motivates us to define the Zernike space as a tensor.}
    \label{fig: zernike_space_figure}
\end{figure}

In this subsection, we take a closer look at the structure of the Zernike correlation matrix $\mA$. As defined in \eref{eq: A matrix}, the $(\vx,\vx',i,j)$th entry of the correlation matrix at the location-pair $(\vx,\vx')$ and the Zernike mode-pair $(i,j)$ is $[\mA]_{\vx,\vx',i,j} = \E[a_{\vx,i} \; a_{\vx',j}]$. Thus, $\mA$ is a four-dimensional tensor, with two dimensions allocated to the pair of spatial coordinates (organized through a column-wise stack), and two dimensions allocated to the pair of Zernike modes. One way to visualize the Zernike space is to consider the illustration shown in \fref{fig: zernike_space_figure}. For a fixed pixel location $\vx$, there is a vector $\va_{\vx} = [a_{\vx,1},\ldots,a_{\vx,N}]$ where $n = 1,\ldots,N$ denotes the Zernike mode index. As we move to another pixel location $\vx'$, the vector becomes $\va_{\vx'} = [a_{\vx',1},\ldots,a_{\vx',N}]$.

For a fixed Zernike mode pair $(i,j)$, the correlation is limited to the spatial axis. This will give us the $(i,j)$th slice of the four-dimensional tensor
\begin{equation*}
\mA_{i,j} =
\begin{bmatrix}
    [\mA]_{\vx_1,\vx_1,i,j} & \dots & [\mA]_{\vx_{p},\vx_1,i,j}\\
    \vdots &  \ddots & \vdots \\
     [\mA]_{\vx_{p},\vx_1,i,j} & \dots &  [\mA]_{\vx_{p},\vx_{p},i,j}\\
\end{bmatrix},
\end{equation*}
where $\vx_1,\ldots,\vx_{p}$ are the $p$ coordinates in the grid. The spatial correlation is well structured if for a fixed $(i,j)$ pair --- it is convenient to describe the tensor with respect to only its auto- or cross-correlation functions. Assuming that the Zernike coefficients across the field of view is homogeneous (under the approximation of \cite{Chimitt2020}), the $(\vx,\vx')$th element $[\mA_{i,j}]_{\vx,\vx'}$ will be the function of $\vx-\vx'$ instead of the absolute positions $(\vx,\vx')$. In this case, we can write $[\mA_{i,j}]_{\vx,\vx'}$ as $[\mA_{i,j}]_{\vs}$ where $\vs = (\vx-\vx')/D$ with $D$ being the aperture diameter\footnote{The necessity to normalize the coordinate using $D$ comes from the fact that the geometry of the optical system has a certain impact to the Zernike coefficients. In particular, two aperture diameters $D_1$ and $D_2$ will lead to two different Zernike space because the imaging systems are viewing through different effective slices as a function of aperture size. Therefore, we define a standardized unitless vector corresponding to correlation length in accordance with \cite{Chimitt2020} to be $\vs = \frac{L(\vtheta - \vtheta')}{D} = \frac{\vx - \vx'}{D}$, where $\vtheta, \vtheta'$ are two vectors pointing from the center of the imaging system to points $\vx, \vx'$, respectively.}. For a grid of points $\vx_1,\vx_2,\ldots,\vx_p$, the matrix $\mA_{i,j}$ takes the form
\begin{equation*}
\mA_{i,j}
=
\begin{bmatrix}
    [\mA]_{\vs_0,i,j} & [\mA]_{\vs_1,i,j} & \dots & [\mA]_{\vs_p,i,j}\\
    [\mA]_{\vs_1,i,j} & [\mA]_{\vs_0,i,j} & \dots & [\mA]_{\vs_{p-1,i,j}}\\
    \vdots &  \vdots & \ddots & \vdots \\
    [\mA]_{\vs_p,i,j} & [\mA]_{\vs_{p-1,i,j}} & \dots & [\mA]_{\vs_0,i,j}\\
\end{bmatrix}.
\end{equation*}
Because of homogeneity, $\mA_{i,j}$ can be decomposed via the Fourier transform.

For a fixed location pair $(\vx,\vx')$, we can obtain another slice of the correlation matrix
\begin{equation*}
\mA_{\vx,\vx'} = \mA_{\vs} =
\begin{bmatrix}
    [\mA]_{\vs,1,1} & \dots & [\mA]_{\vs,1,N}\\
    \vdots &  \ddots & \vdots \\
    [\mA]_{\vs,N,1} & \dots &  [\mA]_{\vs,N,N}\\
\end{bmatrix}.
\end{equation*}
The special case where $\vs = \vec{0}$, which gives matrix $\mA_{\vec{0}}$ is exactly the Noll matrix \cite{Noll_1976} that specifies the correlation between the Zernike modes at a single pixel location $\vx$. In general, the matrix $\mA_{\vs}$ does not have the circulant structure and so it cannot be decomposed via the Fourier transform.

The prior notation allows us to easily identify the homogeneous and the non-homogeneous components of the tensor. The entire Zernike correlation tensor $\mA$ can be written as
\begin{equation*}
\mA =
\begin{bmatrix}
    \mA_{1,1} &\cdots &\mA_{1,N}\\
    \vdots &\ddots &\vdots\\
    \mA_{N,1} &\cdots &\mA_{N,N}
\end{bmatrix}
= \begin{bmatrix}
    \mA_{\vs_0} &\cdots &\mA_{\vs_p}\\
    \vdots &\ddots &\vdots\\
    \mA_{\vs_p} &\cdots &\mA_{\vs_0}
\end{bmatrix},
\end{equation*}
The entire tensor structure is non-homogeneous. Thus, it is not possible to draw samples via the FFT method, which highlights a fundamental limitation of \cite{Chimitt2020} and \cite{Mao_2021_ICCV}.

\subsection{Generation of the Random Zernike Fields}
\label{sec: simulation_fields}

Due to the non-homogeneity of $\mA$, we propose an approximate tensor $\widetilde{\mA}$ which we claim captures a majority of the statistical behavior while having the property of homogeneity. Reserving the accuracy of this approximation for section \ref{sec: justify_approx}, we propose the following method for the generation of the Zernike fields:
\begin{enumerate}
    \item Generate $i=\{1,2,\ldots, N\}$ unit-variance, spatially correlated random fields according to $\mA_{i,i}$. Note this generation uses only \emph{auto}covariance functions. At this stage, our $N$ fields are independent, thus utilizing the FFT-based method based on their homogeneity.
    \item Perform a point-wise mixing of the random fields according to the Noll matrix $\mA_{\vec{0}}$. This mixing is done pixel-wise (across the coefficient index dimension) per pixel.
\end{enumerate}
This generation process will give us random fields which are in accordance with the autocovariance functions, however, for the cross-covariance terms there will exist some deviation. This is most simply presented by the form of the covariance structure of the tensor,
\begin{equation}
    \widetilde{\mA} = \mL\begin{bmatrix}
    \mA_{1,1} & 0 & \dots & 0 \\
    0 & \mA_{2,2} & \dots & 0 \\
    \vdots & \vdots & \ddots & \vdots \\
    0 & 0 & \cdots & \mA_{N,N}
    \end{bmatrix}
    \mL^T,
\end{equation}
where $\mL \mL^T = \mA_{\vec{0}}$. We note the resultant matrix is no longer diagonal. Therefore, the off-diagonal entries of $\widetilde{\mA}$ differ from $\mA$, which will be the focus of our numerical analysis.

\begin{figure}
    \centering
    \includegraphics[width=0.95\linewidth]{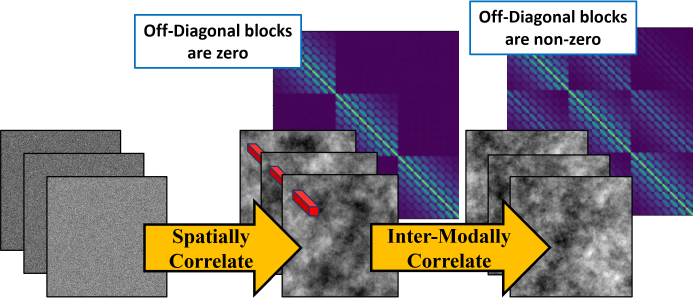}
    \caption{We give a visual representation of the covariance structure as it changes through the generation process. First, white noise has a covariance structure applied to it, resulting in spatially correlated, yet independent random fields. Next, along the index axis, a mixing matrix is applied via Cholesky decomposition. Finally, the resulting fields are correlated both spatially and along the index dimension and have the covariance structure shown.}
    \label{fig: rand_gen}
\end{figure}

We provide a visualization of the resulting covariance matrix in \fref{fig: rand_gen} using a simplistic example in the case of an $8\times8$ image with 3 coefficient fields. Initially, white noise is generated for an $8\times8\times3$ random volume, which is then spatially correlated for each slice according to its autocovariance function. The resulting covariance matrix has a considerable amount of zero-entries corresponding to the other $8\times8$ random fields. After this, the fields are mixed according to the $3\times3$ covariance matrix along the final index axis, after which the covariance matrix becomes much denser. This same principle is extended to the case of an $W\times H$ image with $N$ coefficient fields.

\subsection{The Impact of Cross-Correlation Functions}
\label{sec: justify_approx}
The core innovation in this work is the ability to utilize property of homogeneity for individual fields $a_{\vs,i}$ (or equivalently, $\widetilde{a}_{\vs,i}$). This allows us to quickly draw samples using FFT-based generation followed by a mixing matrix. To justify this approximation, we begin with the following considerations:
\begin{enumerate}
    \item The correlation functions $\mA_{i,j}$ for $i \neq j$ sharply decay and approximately vanish for $s>4$;
    \item The off-diagonal terms for the matrix $\mA_{\vx,\vx}$ (corresponding to indices $i \neq j$) contain a small proportion of the energy.
\end{enumerate}
Together, these two properties suggest that the overall energy lost by the removal of the cross-correlation functions will be negligible. Intuitively, the reason for this lies within the sharp decay of the correlation functions (1), which are already significantly smaller than their counterparts (2).

With this observation, we turn to the numerical analysis performed to more concretely justify this approximation. We wish to measure the energy that will be contained within our approximation and compare it with the energy contained if the entire cross-correlation functions were to be retained. To do so, we integrate outward in the Zernike space and quantify the deviation of $\widetilde{\mA}$ from $\mA$ as a function of $\vs$. While the correlation functions $\mA_{i,j}$ have angular dependencies, we integrate over the angular components for ease of presenting our comparison. We note that along any particular direction individually, the results do not vary considerably. Mathematically, we write the energy contained with all functions considered via the Frobenius norm as
\begin{equation}
    E(s) = \int_0^s  \int_0^{2\pi} d\vs' \trace{\mA_{\vs'}^T \mA_{\vs'}},
\end{equation}
where we've integrated spatially outwards up to $s$, as we intend show the accuracy of the approximation as the size of our field in the Zernike space grows. As $s$ becomes large with respect to the spread of the function, $E(s)$ will converge. The energy within our approximation can be written similarly as
\begin{equation}
    \widetilde{E}(s) = \int_0^s  \int_0^{2\pi} d\vs' \trace{\widetilde{\mA}_{\vs'}^T \widetilde{\mA}_{\vs'}}.
\end{equation}
In addition to comparing the energies separately, we may also consider measuring the difference between them. To this end, we propose to measure
\begin{equation}
    E^{(-)}(s) = \int_0^s  \int_0^{2\pi} d\vs' \left|\trace{\mA_{\vs'}^T \mA_{\vs'} - \widetilde{\mA}_{\vs'}^T \widetilde{\mA}_{\vs'}}\right|,
\end{equation}
where we are only interested in the magnitude of the residual, therefore measuring its absolute value.

\begin{figure}
    \centering
    \includegraphics[width=0.95\linewidth]{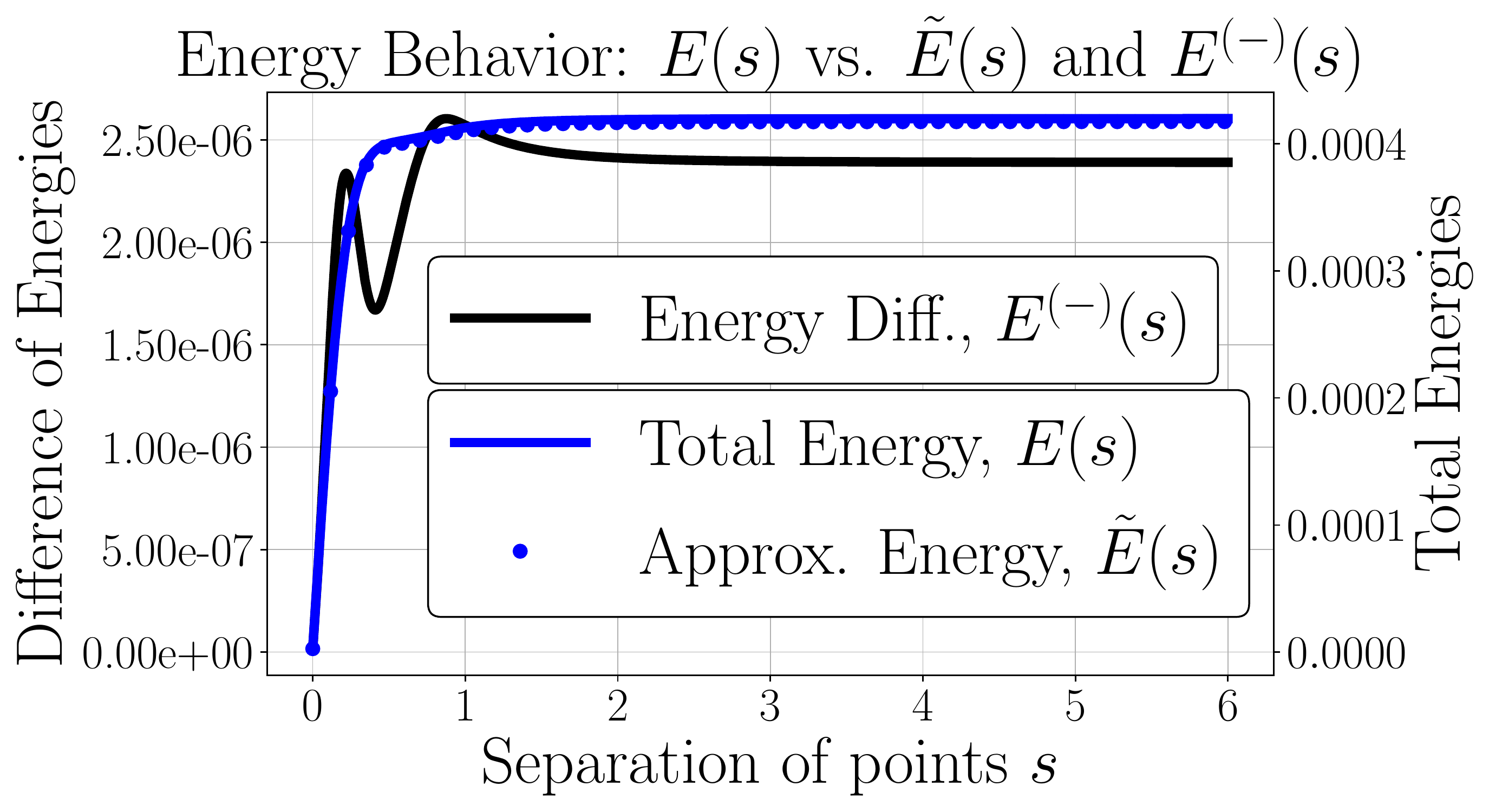}
    \caption{The results from our numerical evaluation of the three proposed energy functions. The axis for $E(s)$ and $\widetilde{E}(s)$ are on the right, while the axis for residual difference energy function $E^{(-)}$ is on the left. We note the apparent match between the two individual energy functions. More importantly, there is a comparably small magnitude for the difference function, suggesting a reasonable match.}
    \label{fig: energy_ratio_fig}
\end{figure}

These three functions will help to provide insight into the accuracy of our approximation. Ideally, we want $E(s)$ and $\widetilde{E}(s)$ to match. However, this alone is not enough to claim accuracy, as this is only a way of measuring the total joint behavior. We additionally then propose to use $E^{(-)}(s)$ to measure the total index-wise residuals. An optimal result for $E^{(-)}(s)$ would be for it to vanish at every point. We present the results of this numerical analysis in \fref{fig: energy_ratio_fig} which we note is cumulative, therefore reaching a steady state means no additional errors will be incurred. We also choose to not include the first three Zernike coefficients (piston and $x$,$y$-tilts) for this plot, as these terms dominate the plot significantly in magnitude, though the conclusion of the analysis is unchanged. With this, we present the following observations:
\begin{enumerate}
    \item At $s=0$ there is no loss in energy/statistical accuracy (e.g. for a single point we are perfectly accurate);
    \item A majority of the errors are for small $\vs$, which is expected as the correlation functions have not yet vanished at this point;
    \item For separations $s > 4$, there is no additional loss in energy and remains at steady state. The energy of the difference $E^{(-)}(s)$ is two orders of magnitude smaller than the total energy, suggesting its overall accuracy.
\end{enumerate}
These three observations suggest that our approximation retains a majority of the field's statistical behavior, while allowing for considerable speed-up.

\subsection{Spatio-Temporally Correlated Fields}

The extension of the proposed ideas to spatio-temporally correlated fields is possible through the adoption of Taylor's frozen flow hypothesis \cite{roggemann1996imaging}. Under this hypothesis, a displacement in time $\vt$ may be written as a spatial displacement $\vy$ through the relationship $v\vt = \vy$, with $v$ as the mean transverse velocity of the turbulent medium. Taylor's frozen flow hypothesis allows us to extend the results by considering $\vz = \vx + \vy$. This will give us a spatio-temporal tensor $\mA_{\vz,\vz'}$ of analogous structure to the previously described $\mA_{\vx,\vx'}$.

A downside of the frozen flow hypothesis is that we need to pre-compute and store multiple three-dimensional random fields. The workaround solution is to enforce the temporal correlation via an auto-regressive process. The auto-regressive model is a low-order approximation. While its theoretical justification is limited, the speed of simulation is marginally impacted. We observe the visual effects under the auto-regressive approximation to be suitable in practice.

\section{Experiments}
In this section, we discuss comparisons between the proposed DF-P2S scheme and the existing approach of split-step. We view the P2S method \cite{Mao_2021_ICCV} as a fast way to generate our PSFs, which may be optionally replaced by \eqref{eq: PSF_formation} at the cost of speed. Therefore, we do not compare directly to this method. We begin with visual comparisons with split-step, which we emphasize are helpful as a reality check, though are not as useful as a statistical comparison. As a result, we take care in comparing our simulation results to the theoretically predicted statistics, which we find to be far more informative. Finally, we compare the runtime and resolution of the DF-P2S approach compared against other methods.

\subsection{Visual Comparisons}
\begin{figure}[ht]
\centering
\includegraphics[width=\linewidth]{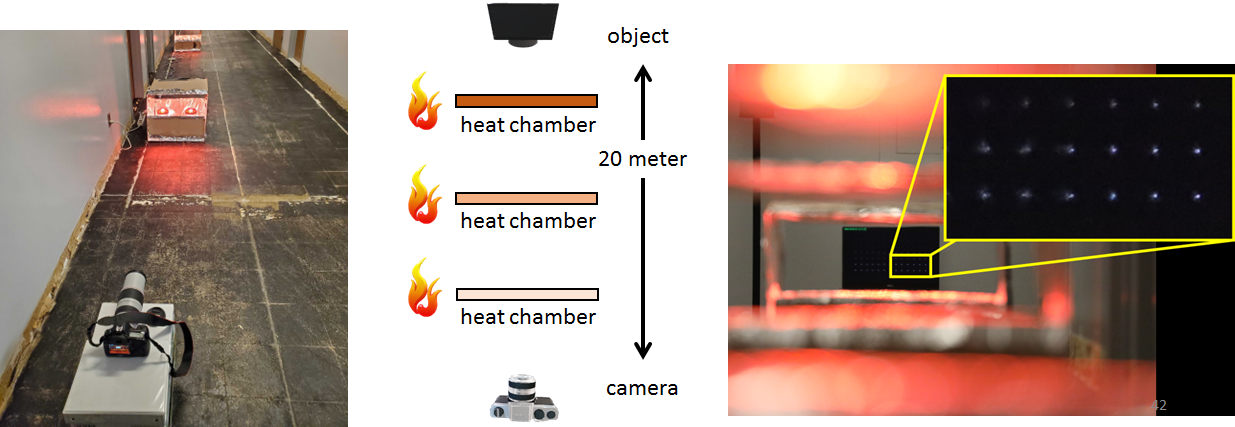}
\caption{A heat chamber is built to collect real turbulence data. The heat chamber consists of three blocks where the total optical path length is 20 meters. The temperature along the path can be controlled by the number of heat lamps that are switched on. At one end of the heat chamber, we put a monitor displaying the target pattern. An 800mm lens camera is placed at the other end of the setup to capture the image.}
\label{fig: heat chamber}
\end{figure}

Visual comparisons require a ground truth image to simulate, as well as turbulent pairs. For this purpose, we use a heat chamber, where we have a series of heating lamps along a 20-meter path, as shown in \fref{fig: heat chamber}. This allows us to easily use a digital image displayed on a monitor at the end of the path.

\begin{figure*}
    \centering
    \begin{tabular}{ccc}
    Heat Chamber (Real Data) & DF-P2S ($<$1 second) & Split-Step ($\approx 20$ minutes, estimated)\\
    \includegraphics[width=0.3\linewidth]{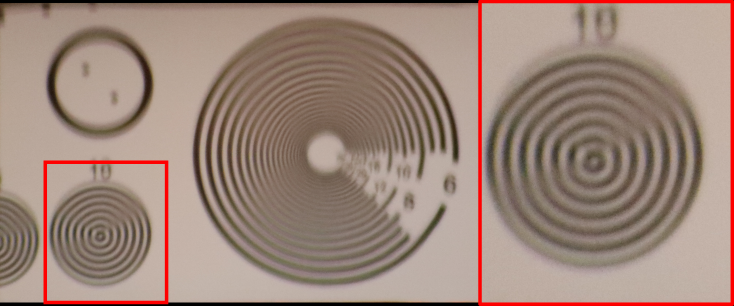} &
    \includegraphics[width=0.3\linewidth]{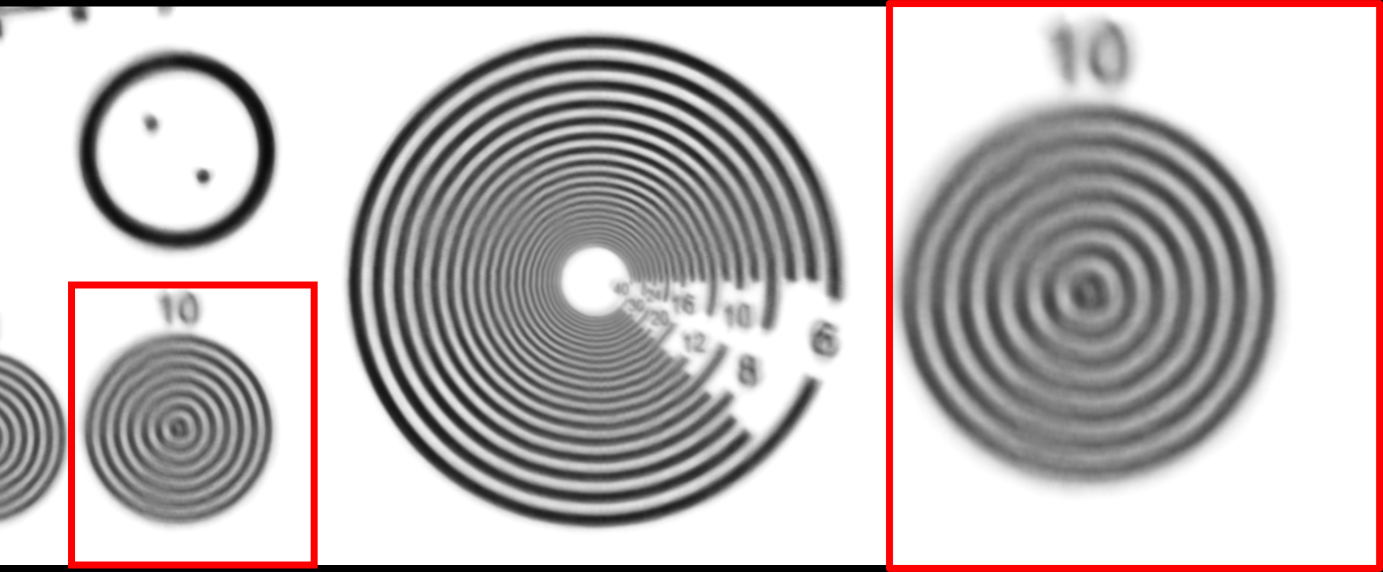} &
    \includegraphics[width=0.3\linewidth]{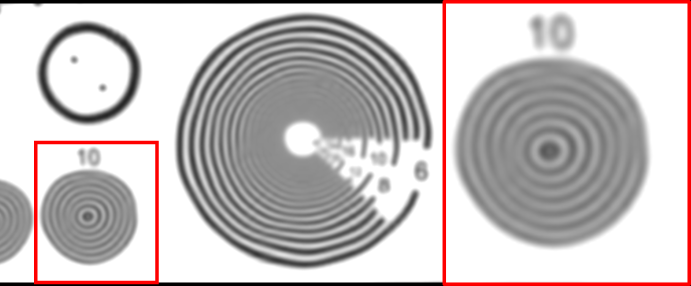} \\

    \includegraphics[width=0.3\linewidth]{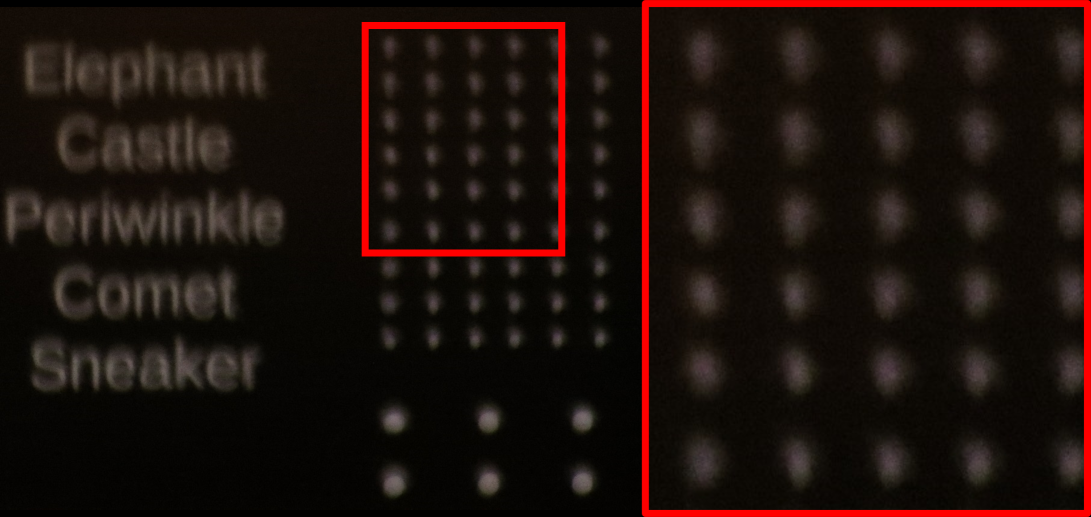} &
    \includegraphics[width=0.3\linewidth]{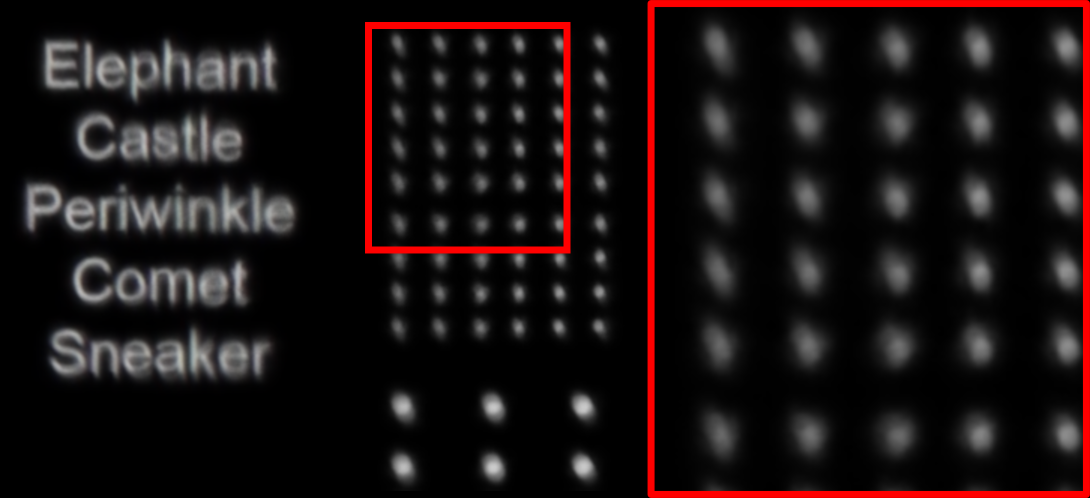} &
    \includegraphics[width=0.3\linewidth]{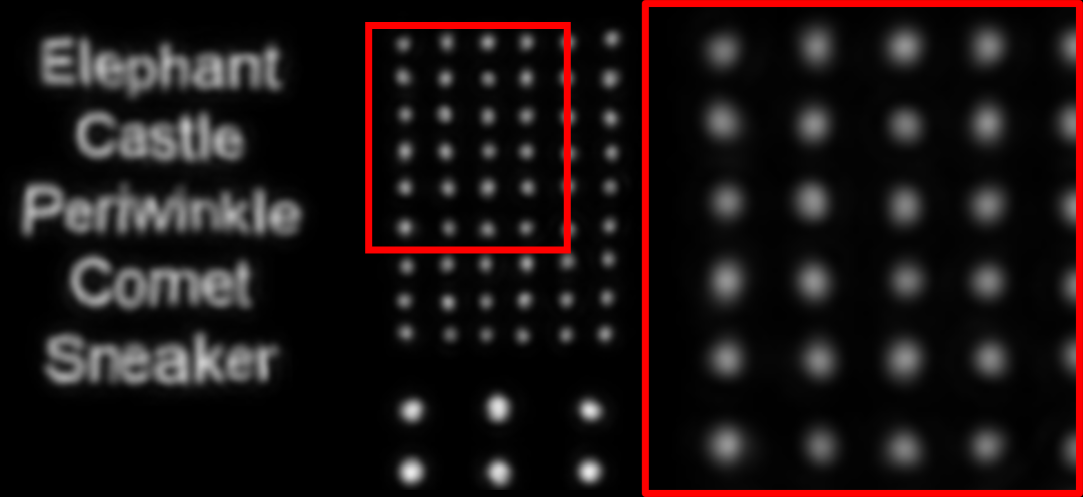} \\

    \includegraphics[width=0.3\linewidth]{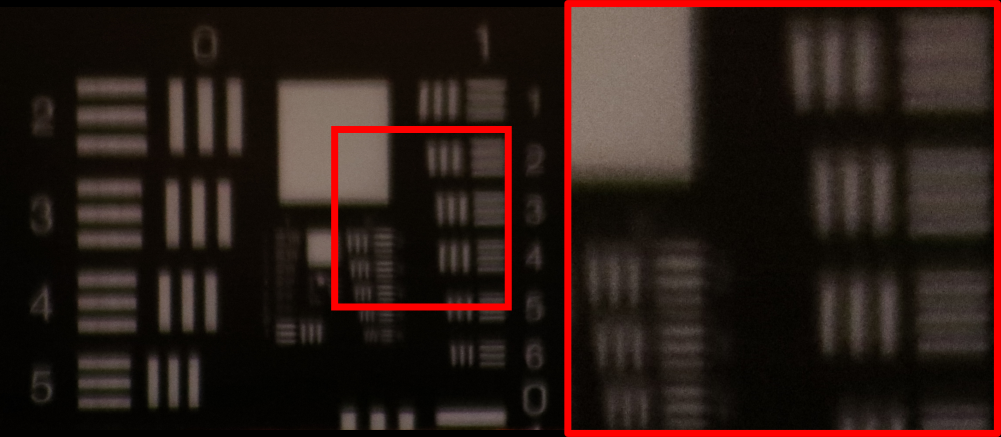} &
    \includegraphics[width=0.3\linewidth]{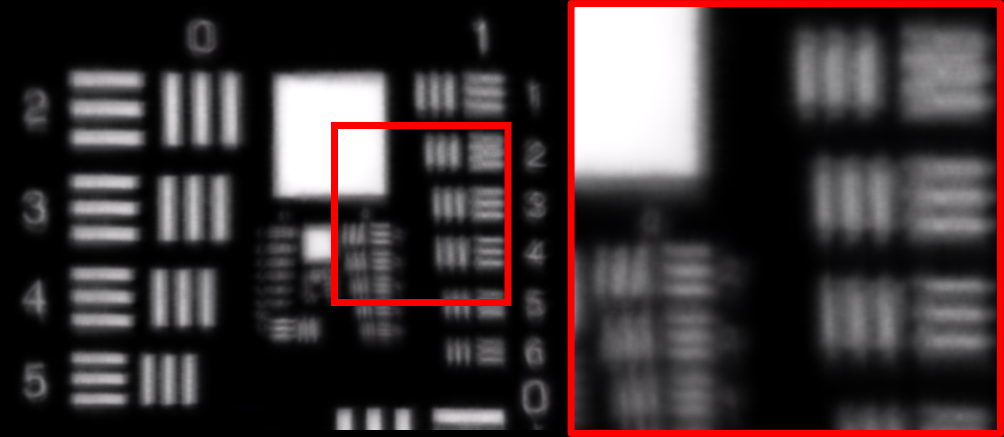} &
    \includegraphics[width=0.3\linewidth]{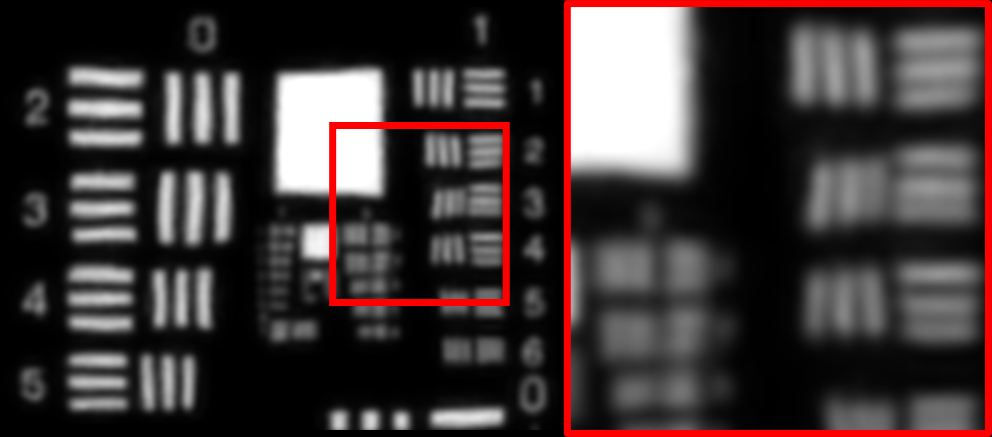} \\

    \end{tabular}
    \caption{Comparison of our simulation approach against our observed heat chamber data and split-step. We observe a reasonable match for both cases, with our approach having the capability to easily operate on high-resolution images. Taking the first row as our example, DF-P2S can operate under a second while split-step would take an estimated time of 20 minutes. Our estimate is based on the grid size used by the split-step.}
    \label{fig: visual_comparisons}
\end{figure*}

We compare our method with our heat chamber data, as well as split-step \cite{Hardie2017, SchmidtTurbBook}. We present a few examples of our visual comparisons in \fref{fig: visual_comparisons}. Visually, we can observe a reasonable match between both generation methods, though we have upscaled the distortions of split-step according to \cite{Hardie2017} to match the image resolution.

The direct visual comparison is a useful reality check; it serves to show that a method can produce visually realistic images. However, there is no easily quantifiable approach to validating a simulator via visual comparison. This motivates us to emphasize other modes of validation.

\subsection{Statistical Validations}
In analyzing the simulation approach, adherence to the desired statistical behavior is a key metric by which we judge the quality of our simulation approach. The validation here is divided into two categories: (1) aperture statistics and (2) spatial statistics. For statistics on the aperture, the key comparison is our generated statistics plotted against the theoretical structure function. With respect to spatial statistics, our comparison is with the known tilt statistics; there is a limitation on the known behavior of the spatial statistics with respect to the blur without the approximation in \cite{Chimitt2020}, so no direct comparison is possible.

\subsubsection*{Aperture Statistics}
\begin{figure}[th]
    \centering
    \includegraphics[width=0.95\linewidth]{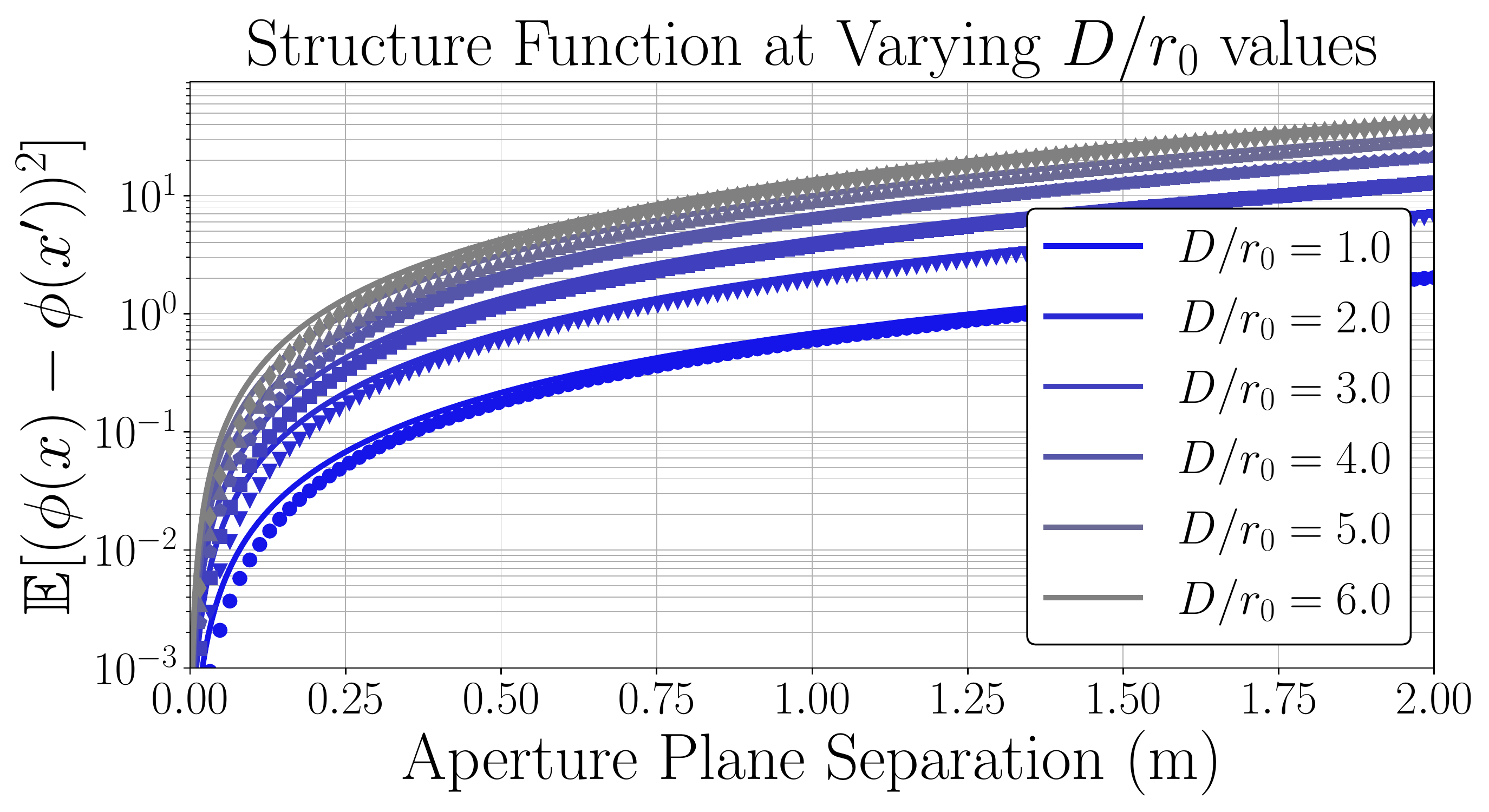}
    \caption{A comparison between the theoretical structure function (solid curves) at different distortion levels vs. our generated statistics (dotted lines). We observe a reasonable match across multiple levels of distortions.}
    \label{fig: struct_fun}
\end{figure}

For the evaluation of aperture statistics, the most important function to match is that of the structure function given by \eqref{eq: fried_struct}. By matching this function, as we show in \fref{fig: struct_fun}, we are able to match any statistical value that can be written in terms of the structure function, such as the Fried parameter. We show a good match to the structure function at varying levels of turbulent distortions.

\begin{figure}[th]
    \centering
    \begin{tabular}{cc}
        \includegraphics[width=0.95\linewidth]{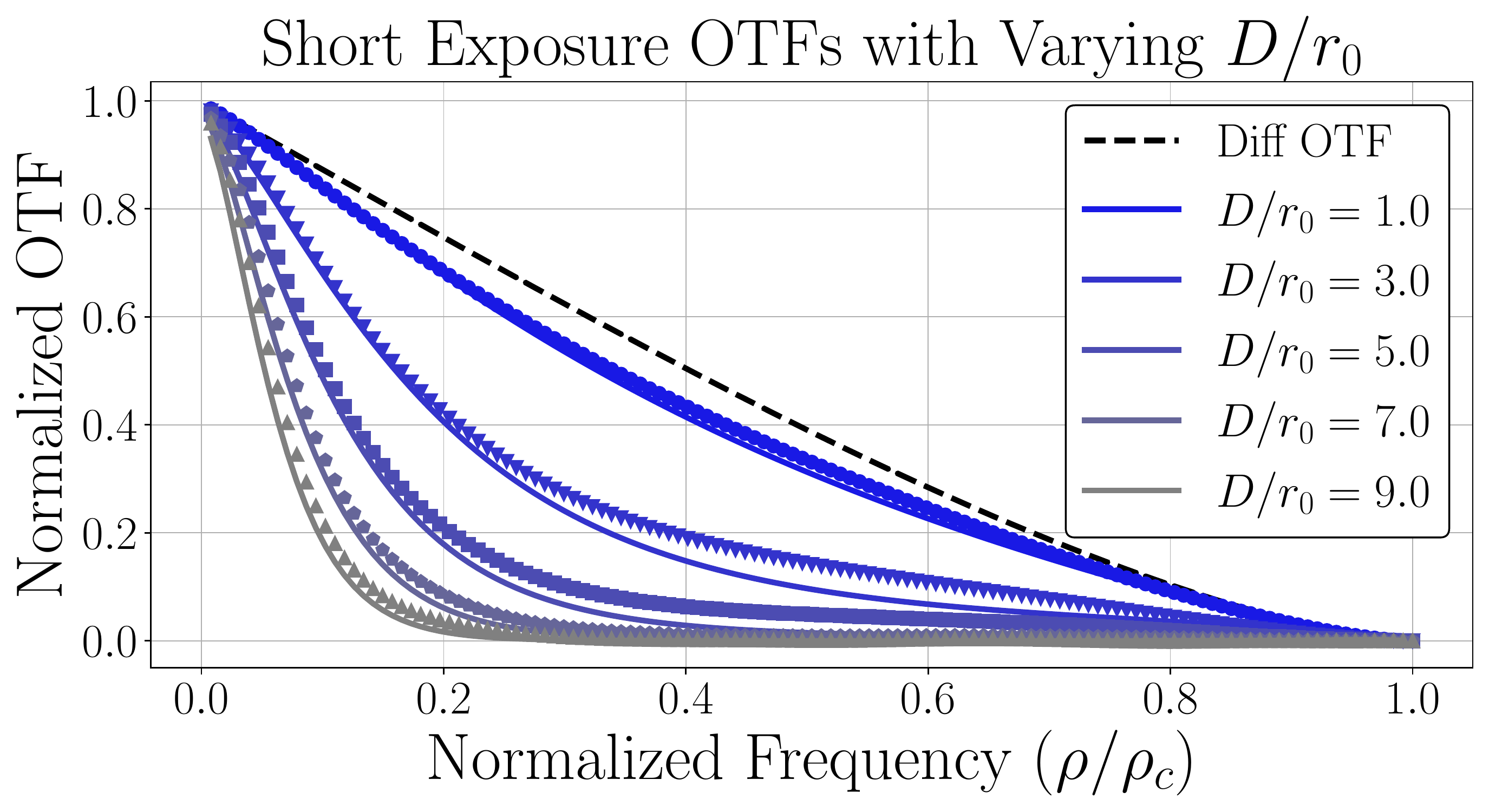} \\
        (a) Short-Exposure OTF\\
        \includegraphics[width=0.95\linewidth]{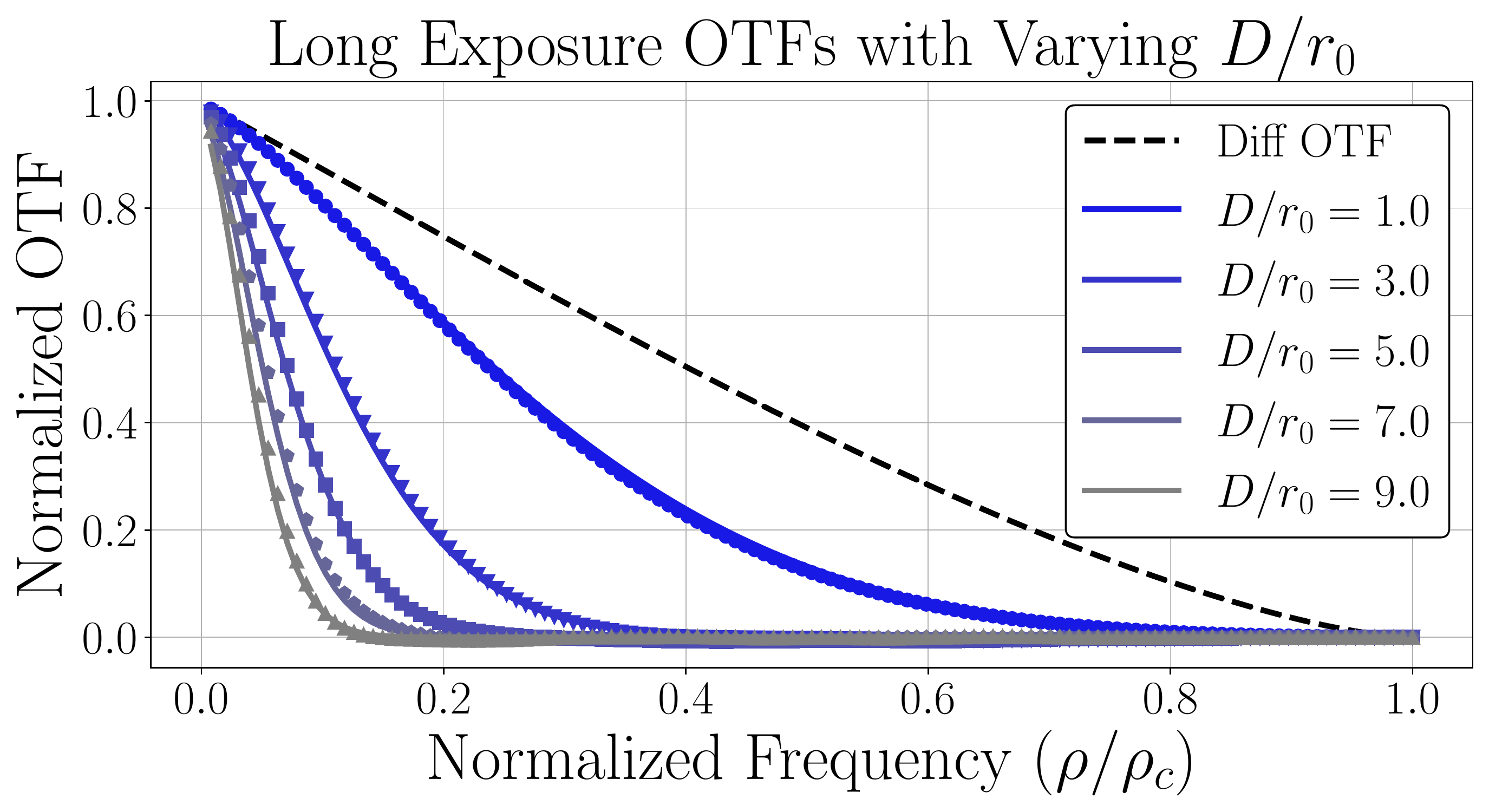} \\
        (b) Long-Exposure OTF
    \end{tabular}
    \caption{Comparing the long and short exposure OTFs at different distortion levels with the theoretical curves (solid lines) and our statistics (dotted lines). }
    \label{fig: SELE}
\end{figure}

We also perform another experiment on the temporal averages of the distortions, the short exposure (SE) and long exposure (LE) optical transfer functions (OTFs). The OTF in general is defined as
\begin{equation}
    H(\vxi) = \left( P(\vxi) e^{-j \phi(\vxi)} \right) \circledast \left( P(-\vxi) e^{j \phi(-\vxi)} \right),
\end{equation}
which is the autocorrelation operation of the overall pupil function and phase distortion. The LE OTF is then given by a temporal average over realizations of individual OTFs. Mathematically, the LE OTF is given as
\begin{equation}
    H_{\text{LE}}(\vxi) = \E \left[ \left( P(\vxi) e^{-j \phi(\vxi)} \right) \circledast \left( P(-\vxi) e^{j \phi(-\vxi)} \right) \right].
\end{equation}
The SE OTF is similarly defined using the temporal average, however, this measure uses a ``tilt-corrected'' phase function $\varphi$ given by
\begin{equation}
    \varphi(\vxi) = \phi(\vxi) - \valpha^T \vxi,
\end{equation}
where $\valpha^T \vf$ is the plane of best fit, effectively removing the tilt. The remaining phase distortion $\varphi$ will then only describe the high-order distortions. This differs from the LE OTF, which includes both blur and tilt/shifting. The SE is then described as
\begin{equation}
    H_{\text{SE}}(\vxi) = \E \left[ \left( P(\vxi) e^{-j \varphi(\vxi)} \right) \circledast \left( P(-\vxi) e^{j \varphi(-\vxi)} \right) \right].
\end{equation}
The LE and SE OTFs have analytic expressions \cite{roggemann1996imaging}, which we compare with the results of our simulation in \fref{fig: SELE} which also includes the diffraction OTF. We observe a wide variety of $D/r_0$ levels that we perform with sufficient accuracy. Some slight deviation may be improved by subharmonic methods such as those described by Schmidt \cite{SchmidtTurbBook}.

\subsubsection*{Spatial Statistics}
\begin{figure}[th]
    \centering
    \includegraphics[width=0.95\linewidth]{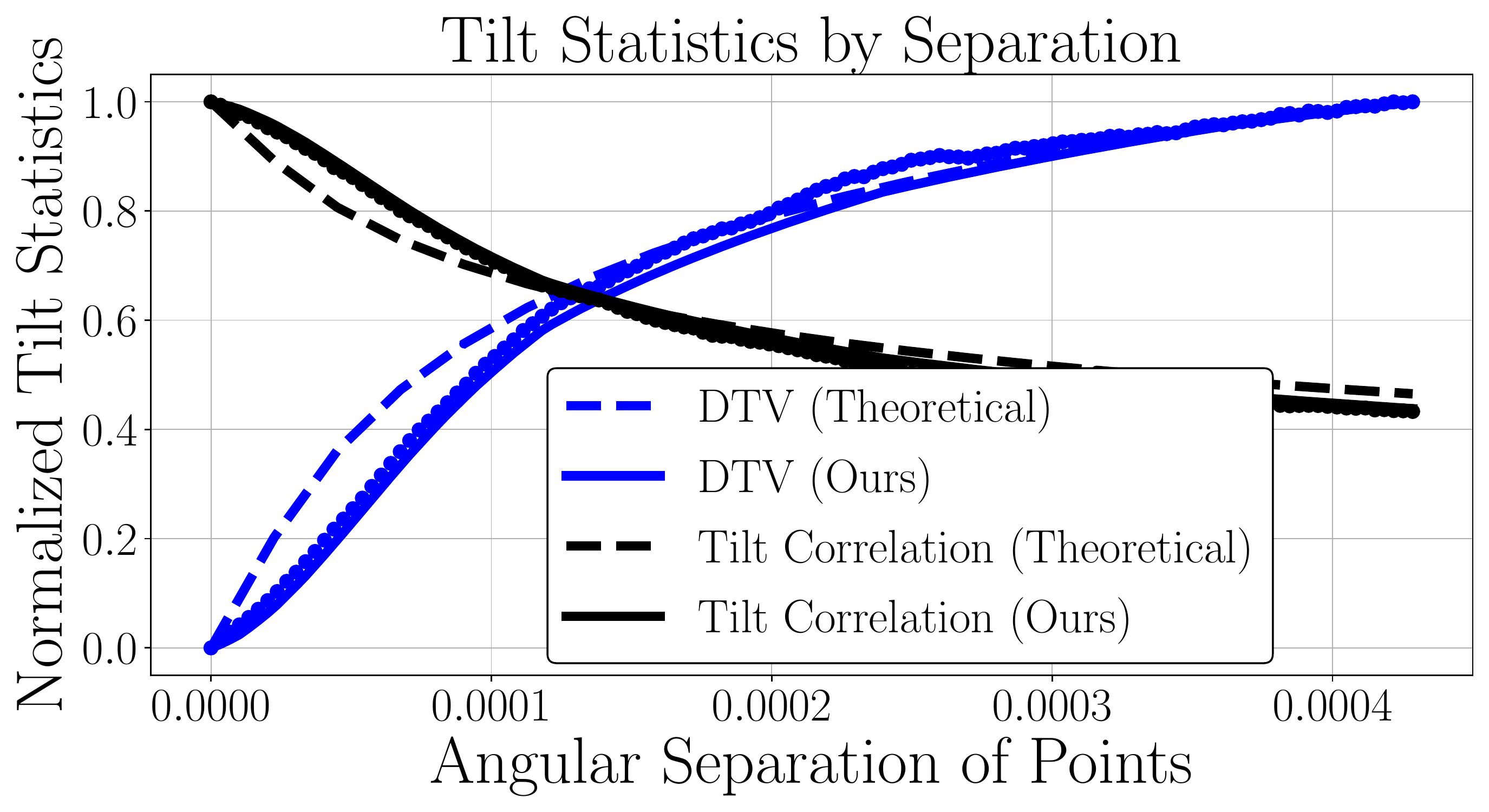}
    \caption{Comparing theoretical vs. empirical spatial tilt statistics. We note that this is for one configuration of camera and geometry, though among our tests we can match to this level of accuracy in a wide variety of scenarios. The limitation of this match is inherent to the approximation of \cite{Chimitt2020}.}
    \label{fig: tilt_dtv}
\end{figure}

For further validation, we also make a comparison with spatial statistics, of which we choose the tilt correlation and differential tilt variance. The tilt correlation measures the similarity between tilts as a function of position, therefore monotonically decreasing, while the differential tilt variance measures the difference between the tilt, and as a result, monotonically increases. The expression for tilt correlation has been carried out by Basu (now Bose-Pillai) et al. \cite{Basu_2015}, which is what we use for our comparison. However, as the expression is analytically cumbersome, we state the definition as given by Fried \cite{Fried1976_variety}
\begin{equation}
    \E [\valpha(0) \valpha(\vtheta)] \propto \iint d\vr d\vr' W(\vr) W(\vr') \vr \cdot \vr' \calD_{\phi}(\vr - \vr', \vtheta)
    \label{eq: basu_angle}
\end{equation}
with a somewhat modified structure function from \cite{Fried1976_variety, Basu_2015}. We present a comparison with these theoretical statistics with our generated statistics in \fref{fig: tilt_dtv}. In addition to tilt correlations, we may also calculate differential tilt variance (DTV) from the tilt correlation via
\begin{equation}
    \E[ (\valpha(0) - \valpha(\vtheta))^2] = 2\left(\E[\valpha^2(0)] - \E[\valpha(0) \valpha(\vtheta)] \right),
\end{equation}
which quantifies the overall variance in the distortions as a function of position of the Zernike tilt terms. We again observe a match across a wide variety of $D/r_0$ values. The deviation of these results is a result of a Taylor series performed in the analysis of \cite{Chimitt2020} upon the structure function. The tilt correlation and DTV changes with different combinations of imaging geometries and camera parameters, though in our tests we can typically match to this degree of accuracy assuming a constant $C_n^2$. For more details regarding this approximation, we would refer the reader to \cite{Chimitt2020}.

\subsection{Speed and Resolution Comparisons}
\begin{figure}
    \centering
    \begin{tabular}{cc}
        \includegraphics[width=0.95\linewidth]{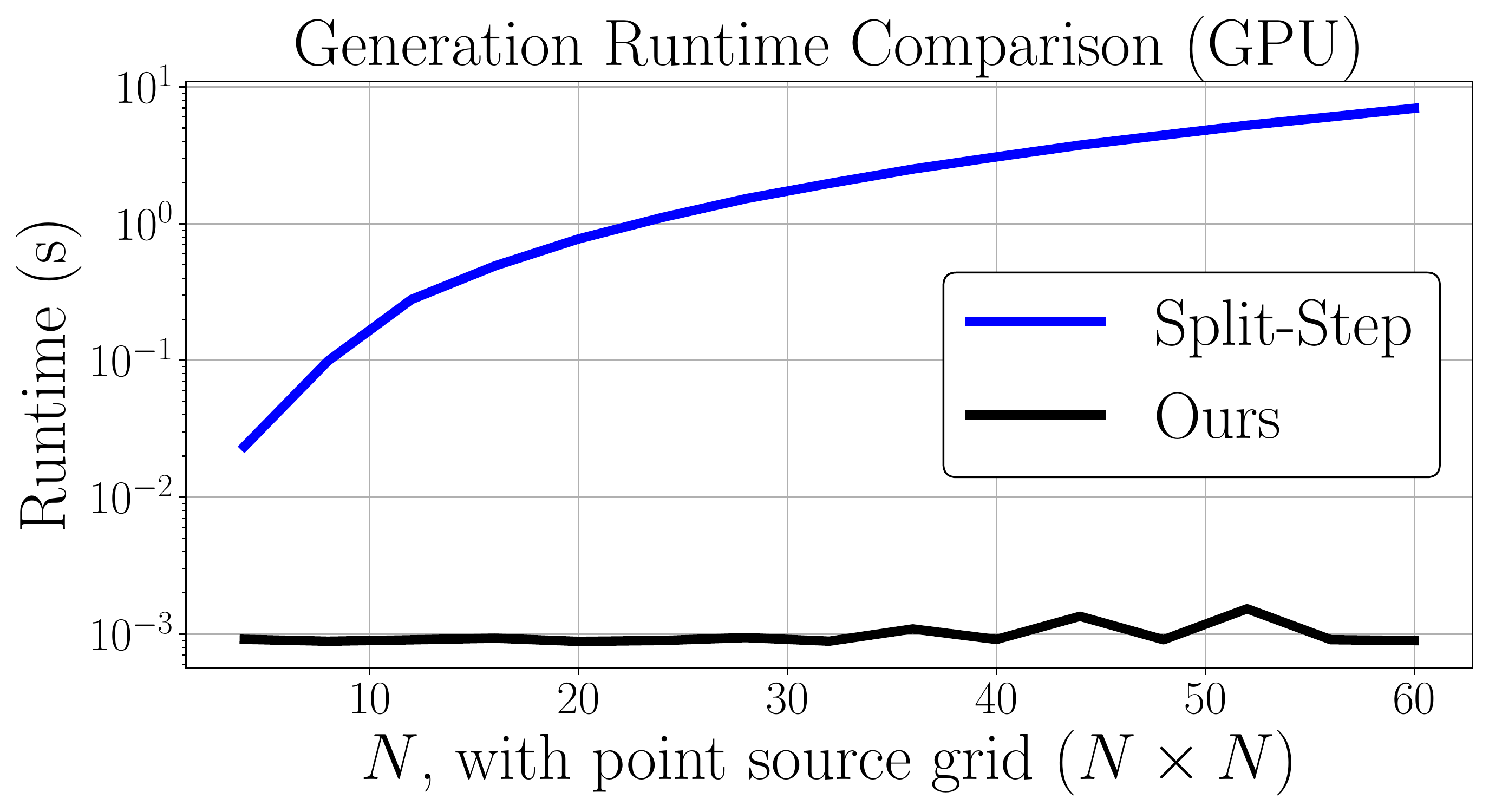} \\
        (a) Low resolution comparison\\
        \includegraphics[width=0.95\linewidth]{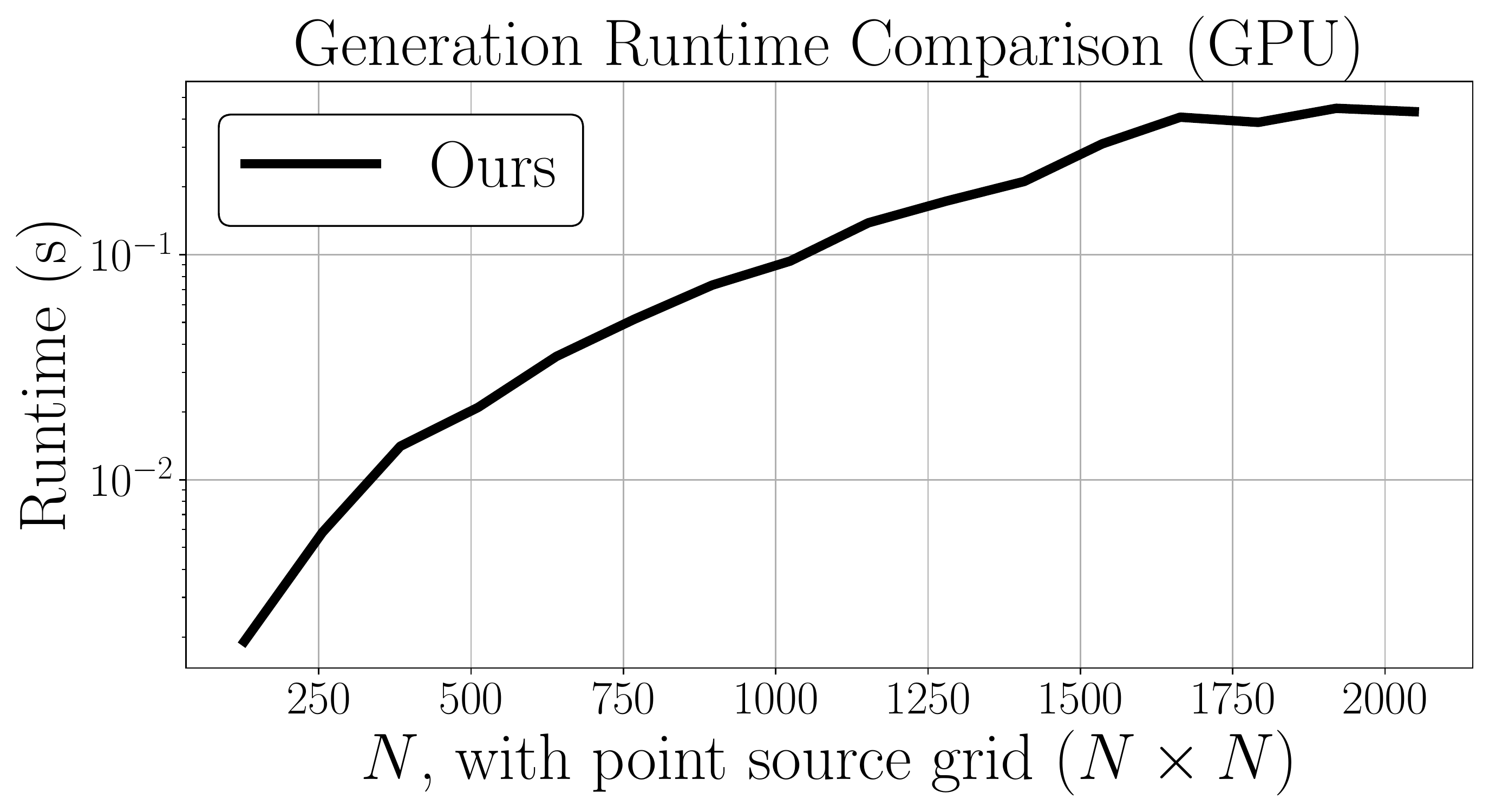} \\
        (b) High resolution comparison
    \end{tabular}
    \caption{A comparison of the runtimes of split-step \cite{Hardie2017} and ours. (a) Low resolution. (b) High resolution. Note that split-step cannot handle any high resolution.}
    \label{fig: runtime_figure}
\end{figure}

For our runtime comparison, we consider the generation contained in \fref{fig: split_vs_multi}. That is, we want to compare the time to generate only the phase distortions, not the application of the point spread functions. We choose not to include this as this can be replaced in either approach with either the analytic formula \eqref{eq: PSF_formation} or the P2S network. We feel this is the most fair comparison of an atmospheric turbulent simulation tool, as the core goal is to produce the phase distortions. Therefore, \fref{fig: runtime_figure} (a) reflects the time to generate the turbulent phase distortions at varying resolutions between our approach and split step. We then show additional resolution for our simulation in \fref{fig: runtime_figure} (b) to which split-step is not scalable. Notably, our simulator can generate an image at an image resolution of $512\times 512$ in the same time as approximately a $10\times 10$ image with split-step, which would then typically be upscaled $4\times$, resulting in a $40\times40$ image.

\subsection{GUI for Real-Time Simulation}
\begin{figure}[ht]
    \centering
    \includegraphics[width=\linewidth]{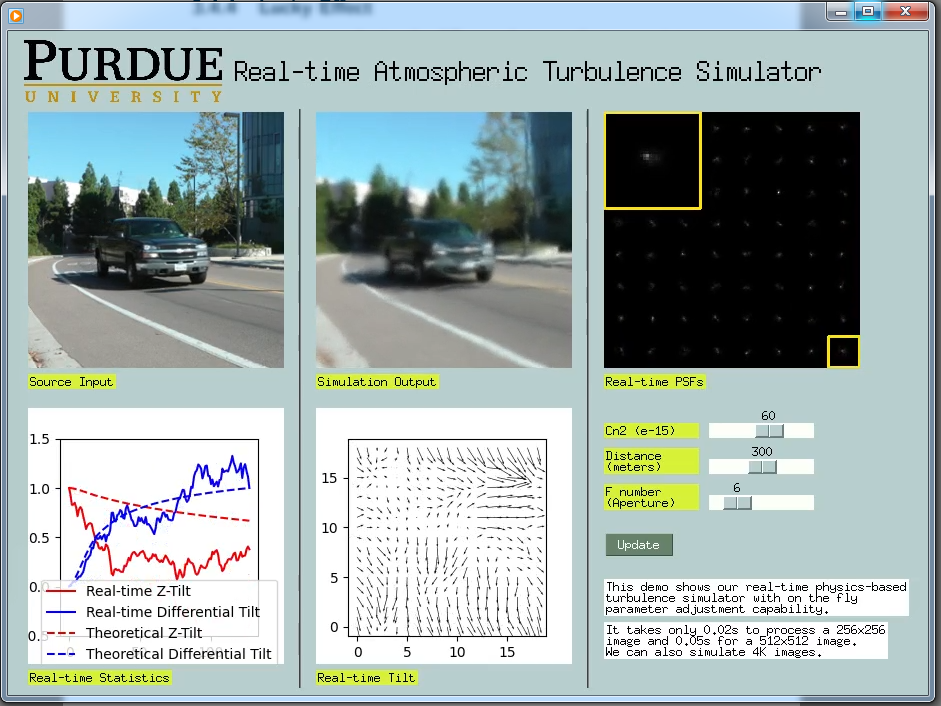}
    \caption{Graphics User Interface (GUI). Our GUI can be connected to a standard camera and generate dense-field turbulence effects at 7 fps for a $512\times512$ image. }
    \label{fig: webcam_demo}
\end{figure}

As a result of our simulation being computationally efficient, a real-time generation of turbulent images can be developed. We show a real-time demonstration using a camera and GPU in which we can display the distorted video stream, pixel displacements, and an $8\times8$ grid of sub-sampled PSFs in \fref{fig: webcam_demo}. Given some interfacing with the camera and calculation of statistics in real-time, this is not performed at the same speed as reported in \fref{fig: runtime_figure}. However, we can achieve approximately 8 frames per second (FPS) for a $512\times512$ image on a nVidia GeForce GTX 1080 Ti GPU with the GUI and additional statistical information displayed. To our knowledge, this is the first simulator in the literature that has the capability of performing a simulation with this level of accuracy at this speed, making this the first real-time demo of turbulent imaging simulation. Additionally, we can modify the imaging geometry and turbulence strength in real-time, for which we have tunable knobs in our GUI.

\section{Conclusion}
In this paper, we've proposed the DF-P2S simulator for imaging through atmospheric turbulence, which we believe to be the fastest simulation modality in the literature. In addition to its speed, it maintains competitive accuracy against with the traditional split-step simulation. This is leveraged by the approximation on the Zernike space covariance tensor, which we justify via numerical experiments. Statistically, the DF-P2S simulation can generate distortions that match with their theoretically predicted curves. We further demonstrate with the speed of our simulator, we can perform these actions in nearly real-time.

\ifCLASSOPTIONcaptionsoff
  \newpage
\fi

\bibliographystyle{IEEEtran}
\bibliography{egbib}

%




\end{document}